\documentclass[11pt,twoside]{article}

\usepackage{asp2006}

\begin{document}
\title{The Extraterrestrial Life debate in different cultures}   
\author{Jean Schneider}   
\affil{LUTh - Paris Observatory}    

\begin{abstract} 
Surprisingly, the question ''Is there Life in the Universe outside  Earth?''
has been raised, in rational terms, almost only in the western literature
throughout the ages. In a first part I justify this statement.
Then I try to develop an explanation of this fact by analyzing the 
different aspects of the notion of ''decentration''.
\end{abstract}

\section{Introduction}
The question ``Are we alone in the Universe?'' is one of the main motivations
of this Conference ``Pathways Towards Habitable Planets.''
It is often claimed to be ``as old as Humanity itself''
It indeed looks very natural since Life is spread out over the whole Earth 
and therefore even a child rising his eyes toward the sky can ask ``Is there also Life out there?'' But, very surprisingly, there is almost no written occurence of this question
in ``non western'' ancient cultures.
In a first part of this paper I justify this statement. Then I will try
to understand why it is so. I will thus be led to first clarify
what can characterize and delimit ``Western'' culture.
Then I will propose a hypothesis to explain why the question
of Life in the Universe has almost never been raised by non-western cultures.
Finally I will address the question ``why did this movement start in Greece?''.

There are generally two ways to consider the question
of extraterrestrial life. First,  the point of view of living organisms, 
leading to the question ``Is there Life elsewhere in the Universe ?'', 
which is the subject of  exobiology and extraterrestrial intelligence, leading to
the question ``Are we alone?'' or``Is there anybody out there?'', 
which is subject of SETI (Search for Extra-Terrestrial Intelligence).
And a different, but connected, question is the nature of Life: how different can it be from terrestrial life? This question is symbolized by the word ``Alien'' often found in the literature. Here I will treat these three questions as if they were only one.

\section{Survey of the world-wide Literature and Traditions}

The question of extraterrestrial life in the literature since
the Greeks has been compiled in the remarkable books ``The Extraterrestrial
Life Debate 1750-1900 - The Idea of a Plurality of Worlds from Kant to Lowell'' (Crowe 1986)
and  and ``The Extraterrestrial life debate, antiquity to 1915''
 (Crowe 2008). They are a must on this topic and  represent an almost exhaustive compilation of all authors having expressed an opinion on this debate. 

It is remarkable that almost all authors entering the debate have expressed 
that the existence of extraterrestrial life seemed natural to them.
Among the most famous authors, the only few remarkable exceptions are 
Aristotles, Augustine, Hegel, Schopenauer and to some
extent Plato. That means a very few skeptics among hundreds of optimists.
 Another curiosity is that, while the debate has increased in intensity among the scientific
community at the end the the XIX$^{th}$ century, almost no philosopher after
Schopenauer was interested in this subject. Only H. Bergson in his {\it Evolution
Cr\'eatrice} and more vaguely C.S. Peirce and W. James did mention
the question of extraterrestrial life. This is strange because several
philosophers at the beginning of the XX$^{th}$ century, like
Husserl, Cassirer, Wittgenstein, were well aware of scientific
developments of their times. To me it remains a mystery. 
It cannot be explained by ignorance: many novelists like Charles Cros, H.G. Wells, G. Flaubert\footnote{in {\it Bouvard and Pecuchet} (Penguin Classics) } A. Strindberg\footnote{In his drama ''Father'', one of the key characters worked on panspermia.},
Marconi, Stapledon\footnote{Olaf Stapledon (1886-1950), a british psychologist, envisaged communication with extraterrestrial in his "Last and First Men" (Stapledon 1930)} and Tristan Bernard\footnote{French humorist, 1866-1947} did contribute to an outreach of the extraterrestrial life debate in the general culture.
Only in the second half of the XX$^{th}$ century
Paul Watzlawick, from the Palo Alto school in sociology, addressed seriously
the question of communication with extraterrestrials (Watzlawick 1976)

The most important, although obvious, observation from Crowe's books is that all authors
cited are Europeans and (after 1800) North-Americans.  No reference
to extraterrestrial life exists in ``Astronomy Across Cultures - The History on Non-Western Culture'' (Selin 2000) nor in ``L'Astronomie des Anciens" (Naz\'e 2009).
There seem to be a few exceptions. I give here the full quotations because they are not very widely known. In the Jewish literature, the ``Guide for the Perplexed'' by Moise Maimonides (circa 1135 - circa 1204)   says
``The whole mankind at present in existence $[$...$]$ and every other species of animals, 
form an infinitesimal portion of the permanent universe $[...]$  it is of great advantage that man should know his station, and not erroneously imagine that the
whole universe exists only for him.'' (Chapter XII p. 268).
But Maimonid was a European Jew living in Cordoba (Spain). He knew well ancient Greeks' 
work and participated in the cultural atmosphere also represented by
Michael Scot (1175 - 1235) and Albertus Magnus (1193 - 1280) for instance 
who were among the Middle Age philosophers supporting the idea of extraterrestrial life. In the Hindu tradition, Bhrigu says in Ch. 9 of the Mahabharata: 

\begin{quotation}
``The sky thou seest above is Infinite. It is the abode of persons crowned with ascetic success and of divine beings. It is delightful, and consists of various regions. Its limits cannot be ascertained. The Sun and the Moon cannot see, above or below, beyond the range of their own rays. There where the rays of the Sun and the Moon cannot reach are luminaries  which are self-effulgent and which possess splendor like that of the Sun or the fire. Know this, O giver of honours, that possessed of far-famed splendor, even these last do not behold the limits of the firmament in consequence of the inaccessibility and infinity of those limits. This Space which the very gods cannot measure is fall of many blazing and self-luminous worlds each above the other. Beyond the limits of land are oceans of water. Beyond water is darkness. Beyond darkness is water again, and beyond the last is fire. Downwards, beyond the nether regions, is water. Beyond water is the region belonging to the great snakes. Beyond that is sky once more, and beyond the sky is water again. Even thus there is water and sky alternately without end.''
\end{quotation}

\vskip -0.05in  
This text, with no clear date of writing (between 400  BC and +400 AD) is interesting, but it is not a rational deduction of extraterrestrial life in the framework of a treatise in Natural Sciences like in Epicure's Letter to Herodotus. In ancient China, there seem to be only two allusions to the plurality of worlds. In the chapter Qianxiang (``The Heavens") of the great encyclopedia {\it Gujin ushu Jicheng} refers to Yi Shizhen who wrote around 1300 AD:

 \begin{quotation}
``Humans and things are without limit, and the same holds for the Earth and the Heavens. As a comparison, when a parasite is in a man's stomach, it does not know that outside this man there are other men; Man being himself in the stomach of the Earth and the Heavens, he does not know that beyond the Earth and the Heavens there are other Earths and other Heavens''.
\end{quotation}

In the book {\it Po Ya Ch'in}, Deng Mu writes around 1000 AD:

\begin{quotation}
 ``Heaven and Earth are large, yet in the whole of empty space they are but as a small grain of rice .... It is as if the whole of empty space were a trunk and heaven and earth were one of its fruits. Empty space is like a kingdom and heaven and earth no more than a single individual person in that kingdom. Upon one tree there are many fruits, and in one kingdom many people. How unreasonable it would be to suppose that besides the heaven and earth which we can see there are no other heavens and no
other earths!"  
\end{quotation}

Let us note here that the latter text by Deng Mu only refers to other ``Earths" and
not to extraterrestrial life. In addition, these seem to be of Buddhist inspiration
and they are very late compared the the IV - III Century BC Greek literature.

To be complete, one must say that there are references to non-human beings in some 
African and Hopi cultures, but they are rather of the ``supernatural" type, like angels for instance. 

In conclusion, the extraterrestrial life debate has by far be initiated
mostly in the arc going from India to Greece and later on in Western Europe.

What is curious is that the question ``Why is this debate
essentially restricted to the western literature?'' has never been discussed,
at least to my knowledge.

\section{Why does the question ``Are We Alone?" exist only in ``Western'' culture ?}
Let us now go beyond a simple historical compilation and try to understand this 
indubitable dis-equilibrium between western and non-western cultures and the roots of
preconditions of the extraterrestrial life debate. Here I will illustrate my argumentation by historical examples. My purpose is nevertheless not a historical perspective. It is rather a-historical and structural. I will develop  a hypothesis which rests on a main guiding principle: ``elsewhere" and ``aliens'' require some distanciation or differentiation.
This principle is schematic compared to the complexity of historical situations, much like how the Galilean inertia principle is apparently contradicted by everyday life dominated by dissipative frictions.

There are two types of distanciation: the distanciation of concepts from
their empirical objects and the spatial distanciation. 
These two aspects are closely connected and in particular spatial distanciation
requires distianciation by concept as a prerequisite.
Let us nevertheless shortly discuss them separately.\\

{\it Distanciation by concepts}

The essence of concepts is to introduce a distance between them and what they are about.
The idea that life can exist \underline{elsewhere} requires that the word "Life''
is not identical with the living beings with which we have personal relationships.
In other words, it requires a \underline{concept} of ``Life". Only concepts can be generalized.
This points toward the ``universalizing" structure of concepts. What is called "abstraction'' is
then the result of this universalization.

We can at this point try to characterize ``Western culture'' as the culture of concepts 
with their mathematization and the logical constraints that they impose.

Concepts are created by the words naming them. This view is illustrated for
instance by the ideas of nominalism (Abelard) and 
the Berkeleysian so-called idealism\footnote{The truth is that the so-called materialism is in
fact a true idealism as we never experience anything like "matter itself", but only perceptions and what language makes of it.}.
Moreover, what is not subject of language cannot be imagined different: to imagine that things are different one must give them names AND detach the word from the designated object. Hence the above-mentioned conceptual distanciation.
An example is given by the idea of ``circle": it is an abstraction insofar as
there is no perfect circle in nature\footnote{See {\it The Origin of Geometry} by E. Husserl.}
and a source of universalization since it allows to put all empirical curves resembling a circle into the same single category. Another, less abstract, example of universalization is given by the introduction of the metrical system which abandoned local customs for a ``universal" length unit (the corresponding "universe" being the Earth, shared by every country). The latter example is a good transition toward spatial distanciation.\\

{\it Spatial distanciation}

Euclid's Elements  introduced a rigorous structure of spatiality, the realm of potential freedom of motion. An important consequence was Thales'  theorem. The latter  permitted one to
make rigorous statements on objects (their length) inaccessible to direct manipulation. As such, it opened the possibility of {\it extra}-polation, the possibility of transferring to distant objects characteristics of objects within our reach, like harboring life
for ``other worlds". It is also worthwhile to note that the idea of proportion underlying Thales'  theorem is in Latin the same word as ``reasoning" (``ratio''), another aspect of the above-mentioned conceptual distanciation. Moreover Euclid's geometry introduced homogeneity of space, opening the possibility that ``here" is not a center, not ``the" center. It is not necessary to recall  the fortune of this idea with the end of geocentrism introduced by Aristarchus of Samos and Copernicus. About the latter, it is interesting to note that there no reference to extraterrestrial life in his writings. In other words, one is thus led from distanciation to decentration.

This homogeneity underlines the great difference with Aristoteles' conception of space for whom the Universe was divided into the Earth (the sublunar world) and Heavens (the superlunar world), by the way like in ancient Chinese astronomy. Both were very heterogeneous and it would have been illogical to transfer to the Heavens something like terrestrial living organisms.

The rationalized structure of space is significantly opposed to the idea of  Yin and Yang where every `yang-like" notion contains some  ``yin-like" quality and vice versa. This Yin-Yang structure is impossible to express in geometrical terms\footnote{It can nevertheless be mathematized in modern terms thanks for instance to "non-well founded" set theory (J. Barwise and J. Etchemendy. ``The Liar" Oxford University Press 1987)   or to Combinatory Logic (Schneider J. 
``La non-tratification'' in {\it La psychanalyse et la r\'eforme de l'entendement} available at
http://www.obspm.fr/$\sim$schneider ).}. This difference in the treatment of space in ancient China and Europe is well illustrated by the difference between Chinese painting and Italian perspective.

To summarize, our hypothesis is that the appearance of the theme of extraterrestrial life in the Indo-European area, and its culmination in Greece, is related to 
the apparition of Euclidean geometry and of the so-called Greek {\it logos}.

\section{Conceptual distanciation, a societal consideration}
In addition to distanciation, another important aspect of concepts is that
they are likely to be shared by every individual. Indeed, it belongs to the essence of concepts that they are not the property of a political power. It results that the political power (King, Emperor) cannot be the source of concepts. They are are their own, impersonal, source. To express it in a radical way, they \underline{are} their own power. In Astronomy, things were very different in ancient China where, for instance, the few astronomical knowledge like the prediction of eclipses or even the calendar were the private property of the Emperor, because they did provide some power. In addition, concepts are open to debate. That is why concepts and democracy go together, if here by democracy one means ``public debate of ideas" rather than such or such election systems, in other words intellectual rather than political democracy. And it is a fact that in this sense democracy has appeared in the part of the world in which  the {\it logos} also appeared.

It is also interesting to note that in the European Age of Enlightenment where the extraterrestrial life debate gained in intensity with authors like Bernard le Bovier de Fontenelle\footnote{{\it Entretiens sur la pluralit\'e des mondes.}}, the idea of decentration gained also a societal tone. This is for instance witnessed by Montesqieu's and Voltaire's work\footnote{Montesiquieu: ``If I knew a thing useful to me but harmful to my family, I would reject it. If I knew a thing useful to my family, but 
useless to my homeland, I would forget about it. If I knew a thing useful to my homeland or to Europe, but prejudicial to the human gender, I would consider it as a crime." in his {\it Carnets}. See also Voltaire's  ``point of view from Sirius" in his {\it Micromegas}.}. There is here a significant contrast with one of the old China's name: "The Empire of the Middle".

One may wonder if such considerations do not lead to a Eurocentrism.
Such a potential Eurocentrism seems to culminate with Kant when he writes in his
{\it Idea for a Universal History from a Cosmopolitan Point of View}:
``.. our continent [Europe] (which will probably give law, eventually, to all the others)...'' 
(9th Thesis)\footnote{He meant ethical laws, pointing toward human rights.}.
 But to this real concern one can reply:

- That it is not an ideological position but a matter of fact that the entire world has
adopted the scientifico-technical concepts.

- That these concepts are not the only respectable values. Ethical values are as important than
scientific rational concepts. And notions like Yin and Yang are more useful in some human affairs than rigid rationality. The German philosopher Martin Heidegger has lengthly developed in his article {\it Dialogue with a Japanese}
 (in {\it On the Way to Language })
that western philosophy has a great deal to learn from the Japanese notion
of {\it koto ba} (which means something like ``gracefulness'').

- In another vein, Greeks' literalism missed the kabbalistic approach of the reading of great texts which  is undoubtedly one of the sources of psychoanalysis.

- One can finally note that if Chinese did have a somewhat elaborated technology, Greeks did
not have a systematic development of  technology. For instance, they 
used steam machines to open the heavy doors of they temples, but
did not think of applying it in a systematic way to everyday practical life
and therefore missed premises of industrialization.

\section{Why did all this start essentially in Greece ?} 
This movement did start in the Indo-European arc (which comprises Arabic countries).
But it exploded in Greece a few centuries B.-C.  One could search for some geographical,
economical or climatic reason for that
\footnote{For instance, the French book {\it Le Secret de l'Occident} (David Cosandey, Editions Arl\'ea, 1997)
claims that science started in Greece because international exchanges were facilitated
by Mediterranean navigation.}. But my thesis is that this Greek
geographical location
is causeless. Its origin is pure genuine fortuitousness, spontaneous generation.
This claim results from the epistemo-analytical view\footnote{Epistemo-analysis
is the psycho-analysis of {\it episteme}, i.e. of knowledge.} according to which
ideas emerge from nowhere. This view is illustrated by the {\it a priori}
essence of concepts pointed out by Kant: concepts do not emerge FROM experience,
they are a prerequisite to make it intelligible. In another domain, 
modern language theories rest
on de Saussure's principle of {\it arbitrariness} of signs: linguistics
symbols are also given {\it a priori} and are not the result of a causal process.

\section{Conclusion}
The personal views presented here are open to debate. Disagreement with
them is of course always possible,but any disagreeing opinion should at least  offer an alternative explanation of the fact pointed out here that the extraterrestrial life debate seems to be essentially restricted to ``Western" literature. This first attempt is not the last word and deserves further investigations, in particular the search for 
the possible occurrence of the extraterrestrial life debate in other parts of the world.

\acknowledgements I am grateful to Anne Cheng, Michel Didier, Subhash Kak, Marc Kalinowski, 
Jean-Claude Marztloff and Tsevi Mazeh for discussions.\\


{\bf References}\\

Crowe, M. 1986, {\it The extraterrestrial Life Debate 1750-1900} (Cambridge University Press

Crowe, M. 2008, {\it The Extraterrestrial Life Debate, antiquity to 1915}
(Cambridge University Press)

Naz\'e, Y. 2009, {\it L'astronomie des anciens} (Belin)

Selin, H. 2000, {\it Astronomy Across Cultures} (Kluwer)

Stapledon, O. 1930, {\it Last and First Men} (Orion Books)

Watzlawick,  P. 1976, {\it How Real is Real} (Random House)





\end{document}